\documentclass[conference,letterpaper]{IEEEtran}
\IEEEoverridecommandlockouts

\usepackage{fancyhdr}
\usepackage[dvipdf]{graphicx}
\usepackage{indent}
\usepackage{amsmath,amssymb}
\usepackage{paralist}
\usepackage{eclbkbox}
\usepackage{subfigure}
\usepackage{multirow}

\makeatletter
\def\tbcaption{\def\@captype{table}\caption}
\def\figcaption{\def\@captype{figure}\caption}

\ifCLASSINFOpdf
\else
\fi

\begin{document}

\title{Emergence of Altruism Behavior for Multi Feeding Areas in Army Ant Social Evolutionary System
\thanks{\copyright 2014 IEEE. Personal use of this material is permitted. Permission from IEEE must be obtained for all other uses, in any current or future media, including reprinting/republishing this material for advertising or promotional purposes, creating new collective works, for resale or redistribution to servers or lists, or reuse of any copyrighted component of this work in other works.}
}

\author{\IEEEauthorblockN{Takumi Ichimura and Takuya Uemoto}
\IEEEauthorblockA{Department of Management and Systems,\\ Prefecture University of Hiroshima,\\ Hiroshima, 734-8558 Japan\\
E-mail: \{ichimura@pu-hiroshima.ac.jp, yslius7221@gmail.com\}}
\and
\IEEEauthorblockN{Akira Hara}
\IEEEauthorblockA{Graduate School of Information Sciences,\\ Hiroshima City University,\\ Hiroshima, 731-3194 Japan\\
E-Mail: ahara@hiroshima-cu.ac.jp}
}

\maketitle

\begin{abstract}
Army ants perform the altruism that an ant sacrifices its own well-being for the benefit of another ants. Army ants build bridges using their own bodies along the path from a food to the nest. We developed the army ant inspired social evolutionary system which can perform the altruism. The system has 2 kinds of ant agents, `Major ant' and `Minor ant' and the ants communicate with each other via pheromones. One ants can recognize them as the signals from the other ants. The pheromones evaporate with the certain ratio and diffused into the space of neighbors stochastically. If the optimal bridge is found, the path through the bridge is the shortest route from foods to the nest. We define the probability for an ant to leave a bridge at a low occupancy condition of ants and propose the constructing method of the optimal route. In this paper, the behaviors of ant under the environment with two or more feeding spots are observed. Some experimental results show the behaviors of great interest with respect to altruism of ants. The description in some computer simulation is reported in this paper.
\end{abstract}



\section{Introduction}
\label{sec:Introduction}
In animal societies, self-organization is the theory of how minimal complexity in the individual can generate greater complexity at the population. The rules specifying the interactions among the components in the system are implemented by using only local information without global information. Deneubourg {\it et al.} \cite{Deneubourg89} developed the model of collective decision making without any form of centralized control. The model was developed to show the characteristic patterns of self-organization by Monte Carlo simulation. In the study of social evolution, army ant performs altruism as one behavior of complexities, where each individual reduces its own fitness but increases the fitness of other individuals in the population. Such behaviors seem to be involved acts of self-sacrifice in order to aid the others. In evolutionary biology, such a behavior is called reciprocal altruism. The concept was initially developed to explain the evolution of cooperation as mutually altruistic acts\cite{Trivers71}. The basic idea is close to the strategy of ``equivalent relation'' in the study of strategic decision making.

Army ants are characterized by their two different phases of activities, a nomadic phase and a stationary phase. During the nomadic phase, army ants move during the day to capture insects, spiders, and so on. The stationary phase starts when the larvae pupate for a few weeks. Moreover, army ants build a living nest with their bodies instead of building a nest like other ants. Each ant will hold on to the other legs and form a linked chain or a ball structure. This behavior is known as a bivouac. This allows the bridging of an empty space. In order to address the self-assembled structure as a particular type of aggregation, Deneubourg {\it et al.} defined the probability of an ant entering or leaving chain in \cite{Deneubourg02}. Moreover, they showed that the gregarious behavior facilitates cooperation by Blattella germanica in shelters during the resting period. The probability to leave the shelter was defined.

Ishiwata {\it et al.} \cite{Ishiwata11} developed the simulation system for the foraging behavior and the altruism of army ants by using Swarm library, {\it Swarm-2.2}\cite{SwarmLib}. (The original website www.swarm.org is in the process of being rebuilt.) The probabilities to form the chain defined in \cite{Lioni01, Lioni04} was used in their simulation experiments. The number of neighboring active ants is considered as the condition for altruistic behavior. Their simulation results show a mimic altruistic behavior.

By inspiring Ishiwata's study, Douzono {\it et al.} developed the multi-agent simulation system to execute more realistic altruistic behavior where two or more kinds of agents realize the sub-tasks of army ants \cite{Douzono12, Ichimura12}. According to the environment in \cite{Douzono12}, the simulation results reported that the optimal path from the food to the nest cannot be always found, because two or more chains in the environment were formed. Although more emergence of altruistic behaviors was observed, but the capabilities of forming chain was dispersed. As a result, the performance of foraging decreases and some ants took a circuitous route. On the contrary, Ichimura {\it et al.} defined the evaporation rate dues to normal distribution probability and the probability to leave from the chain when the ants in its neighbor region depart gradually in \cite{Ichimura12}. The altruism simulation results are reported to find more optimal paths from food to the nest.

In this paper, we observed the behaviors of ant agents under the multi feeding spots in the same environment of \cite{Ichimura12}. Some experiments with different ratio of feed size were investigated. In general, ant agents take an action to be concentrated in the largest feeding spot. The shortest path from the spot to the nest is constructed and the ants bring feed to the nest. Then, the feeding spots will be disappeared in the order of larger spot. However, it has turned out that there is a certain tendency without regard to the size of feed. The altruism behavior does not work well and the bridge will be broken, if enough ant agents are not gathered into the ditch. As a result, the food at the spots remains to the end of simulation. We report the experimental results for the emergence of altruism behavior for multi feeding spots in this paper.

The remainder of this paper is organized as follows. Section \ref{sec:SimulationEnvironment} describes the simulation environment with Swam library. Section \ref{sec:AgentBehaviors} defines the behaviors of agents such as search phase, homing phase (return to the nest), and altruism phase. Section \ref{sec:Proposedmethod} describes the proposed method related to pheromone and the leaving probability from chain. Experimental results for simulations are described in Section \ref{sec:ExperimentalResults}. In Section \ref{sec:Conclusivediscussion}, we give some discussions to conclude this paper.

\section{Simulation Environment}
\label{sec:SimulationEnvironment}
\begin{figure}[tb]
\begin{center}
\includegraphics[scale=0.7]{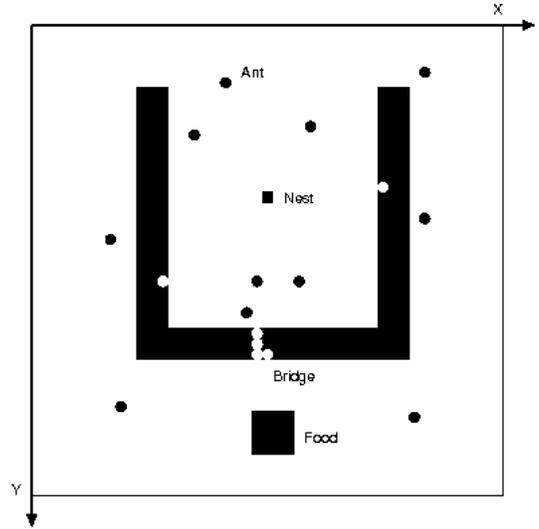}
\caption{Environment in Army Ant Simulation System}
\label{fig:Environment}
\end{center}
\end{figure}

The {\it swarm} is the basic unit of simulation for a collection of agents executing a schedule of actions. The {\it Swarm} provides object oriented libraries of reusable components for building models and analyzing, displaying, and controlling experiments on those models. We executed the altruism simulation system by pheromone evaporation and its diffusion in army ant multi agent systems. The developed system is depicted in $100 \times 100$ 2D space as shown in Fig.\ref{fig:Environment}. The solid-filled rectangle, which consists of 3 kind of bars: `leftbar', `rightbar', and `centerbar', divides the space into 2 parts. The inner part is the nest region and the bottom part under the rectangle is the food source part.
 
The 4 coordinates (x,y) of leftbar, rightbar, and centerbar are \{(30, 30), (33, 30), (30, 70), and (33, 70)\}, \{(70, 30), (73, 30), (70, 70), and (73, 70)\}, and \{(30, 70), (73, 70), (30, 73), and (73, 73)\}, respectively. Each bar represents a ditch and the width of ditch is 3. The center rectangle of the space is `nest' and the bottom rectangle under the bar is `food source.' Since an ant cannot cross the ditch by itself, some ants begin altruistic behavior to cooperate with each other. The two hypotheses were proposed as the judgment criteria for altruistic activity, Model 1: Based on the Presence of Neighboring Ants and Model 2: Based on the Presence of Pheromone \cite{Ishiwata11}. In Model 1, an ant will start formation of living bridge over a gully only when neighboring ants are present. Hypothetically, this approach will be more efficient compared to forming a bridge blindly. In Model 2, the places where pheromone concentrations are higher than a fixed level are the locations that many ants have passed and/or will pass through in future. 

\begin{figure}[!tb]
\begin{center}
\includegraphics[scale=0.8]{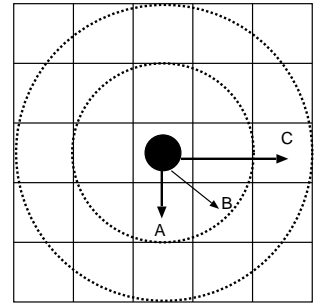}
\caption{The area of activities}
\label{fig:Behaviors}
\end{center}
\end{figure}

Fig.\ref{fig:Behaviors} shows the area of activities and the visual field by an army ant. In this paper, ant agents can move in the diagonal direction, but the scattered pheromone diminishes compared to the adjacent positions on up and down, left and right. The distance from a position to the neighbor is defined $Distance$. For example, the distance to `A' and `C' in Fig.\ref{fig:Behaviors} are 1 $Distance$ and 2 $Distance$, respectively. Practically,`B' is $\sqrt{2} Distance$. However, for the sake of ease, we define `B' as 1 $Distance$ in the diagonal direction.

\section{Agent Behaviors}
\label{sec:AgentBehaviors}
The actions include foraging for foods and transport of them and communications with neighboring ants using pheromone. The pheromone is released by an agent when it finds food. Once the pheromone is attenuated and is dispersed, the information about the food position is disseminated among the ants.

The system has 2 kinds of ant agents, `Major ant' and `Minor ant' and the ants communicate with each other via pheromones. Major ant scatters pheromones and moves throughout in the environment. On the other hand, Minor ant makes a mimic altruistic behavior to foraging and transporting. Douzono {\it et al.} show the numerical superiority in case of the 2 variants of ants \cite{Douzono12}. In this paper, the experimental simulation related to the altruistic behaviors has the 2 kinds of ant agents.

A minor ant agent aims to find a food source and then to return to the nest. If there is a ditch in the path among them, the ant will build a bridge. The 3 kind of states are defined according to the behavior of ants \cite{Ishiwata11}.

\begin{description}
\item[Search State]\mbox{}\\
Search state is an initial condition of agents to seek the food source by random walk. Once an agent reaches the food, it moves into Return state. The ant takes a food on the way back to the food until the food source becomes empty.
\item[Return State]\mbox{}\\
In Return state, an agent comes back to the nest carrying the food. After reaching the nest, the agent moves into Search state.
\item[Altruism State]\mbox{}\\
Some agents stop walking before a ditch and come together as flock. Two or more agents will build a bridge.
\end{description}

\subsection{Search State}
\label{sec:SearchState}
\begin{center}
\begin{indentation}{0cm}{0.5cm}
\begin{breakbox}
\smallskip
\begin{enumerate}
\item Search a food source\\
\label{item:Search:search_food}
First, search a food source within 1 $Distance$. If the agent finds a food at the destination, moving into Return state.
\item Perception of pheromone\\
\label{item:Search:search_perception}
Perceiving pheromone within 2 $Distances$.
\item Search the other agents\\
\label{item:Search:exsistence_otheragents}
Checking the other agents within 2 $Distances$. If the other agents stays, go to \ref{item:Search:search_ditch}). Otherwise, go to \ref{item:Search:randomwalk}).
\item Search a ditch\\
\label{item:Search:search_ditch}
Search a ditch within 1 $Distance$. If there is a ditch, transit to Altruism State. Otherwise, go to \ref{item:Search:randomwalk}).
\item Move\\
\label{item:Search:move}
Move to the other position according to the scattered pheromone described in section \ref{sec:PheromonUpdate}.
\item Random Selection of Walking Direction\\
\label{item:Search:randomwalk}
Check whether the other agent stays or a ditch exists at the next position except going straight ahead. The next position is selected with an arbitrary probability $\alpha$. If the position is empty, go back to \ref{item:Search:search_food}). Otherwise, select another position. Moreover, if the agent is surrounded to other agents or a ditch, it stays at the same position until the neighbor becomes empty.
\end{enumerate}
\end{breakbox}
\end{indentation}
\figcaption{The search algorithm}
\label{fig:Search}
\end{center}

\subsection{Return State}
\label{sec:ReturnState}
\begin{center}
\begin{indentation}{0cm}{0.5cm}
\begin{breakbox}
\smallskip
\begin{enumerate}
\item Current Position\\
\label{item:Return:currentposition}
Check the current position of an agent. If it is in the nest, go to Search State. Otherwise, it goes to next step to move to the nest.
\item Search a ditch\\
\label{item:Return:search_ditch}
If there is a ditch within 1 $Distance$, go to \ref{item:Return:randomwalk}). Otherwise, it moves a next position to the nest and go to \ref{item:Return:currentposition}).
\item Random Walk\\
\label{item:Return:randomwalk}
Check whether the other agent stays or a ditch exists at the randomly selected next position except going straight ahead. If the position is empty, go back to \ref{item:Return:currentposition}). Otherwise, select another position. Moreover, if the agent is surrounded to other agents or a ditch, it stays at the same position until the neighbor becomes empty.
\end{enumerate}
\end{breakbox}
\end{indentation}
\figcaption{The homing algorithm}
\label{fig:Homing}
\end{center}

\subsection{Altruism State}
\label{sec:AltruismState}
\begin{center}
\begin{indentation}{0cm}{0.5cm}
\begin{breakbox}
\smallskip
\begin{enumerate}
\item Search the other agents\\
\label{item:Altruism:exsistence_otheragents}
If there are $n$ agents within 2 $Distances$, the agent stays with an arbitrary probability $1-P_{i}$ described in the section \ref{sec:LeavingChainProbability} and continues to check its surrounded situation. Otherwise, go to \ref{item:Altruism:goto_search_state}) with the probability $P_{i}$.

\item Go to Search State\\
\label{item:Altruism:goto_search_state}
Select a position within 1 $Distance$ in the part of chain. If the position is empty, go to the position, and then make the transition to Search State. Otherwise, the ant is embedded in the chain.
\end{enumerate}
\end{breakbox}
\end{indentation}
\figcaption{The altruism algorithm}
\label{fig:Altruism}
\end{center}

\subsection{Pheromone Update}
\label{sec:PheromonUpdate}
In many works related ant systems, the ants communicate with each other via the pheromone dissemination. However, the researchers have discussed only about the pheromone on the ground. We consider that the pheromone evaporates and spreads into the space in order to take into consideration the influence distributed in the air. The ant in this study can recognize the volatilization of pheromone in the space, but not know the pheromone on the ground. Based on such an idea, pheromone update process is executed by Eq.(\ref{eq:pheromone_update1}) and Eq.(\ref{eq:pheromone_update2}).

\begin{eqnarray}
\nonumber &&space'_{(x,y)}(t)=r_{A} \times space_{(x,y)}(t)\\
\nonumber &&\hspace{8mm}+r_{B} \times (\sum_{p}space_{(i_{p},j_{p})}(t)-4space_{(x,y)}(t))\\
\nonumber &&\hspace{8mm}+r_{C} \times (\sum_{q}space_{(i_{q},j_{q})}(t)-4space_{(x,y)}(t))
\end{eqnarray}
\begin{eqnarray}
\nonumber (x_{i_{p}},y_{j_{p}})&=&\{(x,y+1), (x,y-1), (x+1,y), (x-1,y)\}\\
\nonumber (x_{i_{q}},y_{j_{q}})&=&\{(x+1,y+1), (x+1,y-1),\\
&& (x-1,y+1), (x-1,y-1)\},
\label{eq:pheromone_update1}
\end{eqnarray}
\begin{equation}
space_{(x,y)}(t+1)=space'_{(x,y)}(t)+r_{e}*ground_{(x,y)}(t),
\label{eq:pheromone_update2}
\end{equation}
\begin{equation}
ground_{(x,y)}(t+1)=ground_{(x,y)}(t)-r_{e}*ground_{(x,y)}(t),
\label{eq:pheromone_update3}
\end{equation}
where $space_{(x,y)}$ means the amount of pheromone in the space over the position $(x,y)$ in Eq.(\ref{eq:pheromone_update1}). $r_{A}$ is a decay rate. $r_{B}$ is the diffusion rate in the direction of up and down, left and right. $r_{c}$ is the diffusion rate in the direction of the diagonal. $ground_{x,y}(t)$ means the pheromone amount on the ground at the position $(x,y)$ in Eq.(\ref{eq:pheromone_update2}). $r_{e}$ is the evaporation rate.

\subsection{Multi Feeding Spots}
\begin{figure}[!tb]
\begin{center}
\includegraphics[scale=0.25]{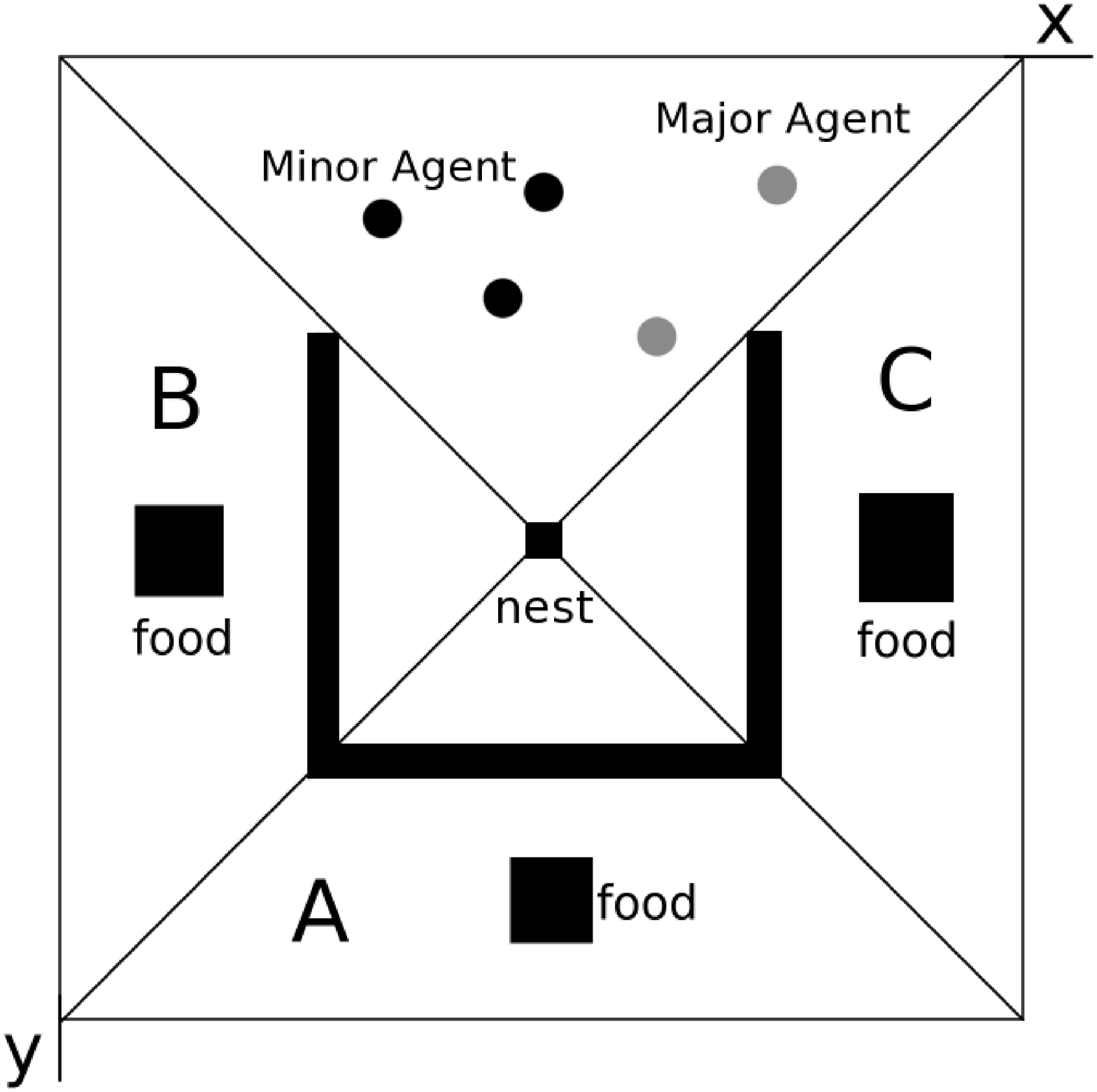}
\caption{The environment for multi feeding spots}
\label{fig:MultiFeeds}
\vspace{-3mm}
\end{center}
\end{figure}

Fig. \ref{fig:MultiFeeds} shows the environment with multi feeding spots to extend the simulation system. As shown in Table \ref{table:feedratio}, we investigate the behaviors of ant for some ratios of food size.

\begin{table}[ht]
\begin{center}
\caption{The size of feeding spots}
\begin{tabular}{l|c|c||l|c|c} \hline
Env. & Agents & A:B:C & Env. & Agents & A:B:C\\ \hline
1-1 & \multirow{4}{*}{\shortstack{Major:3\\Minor:100}} & 2:1:1 &2-1 & \multirow{4}{*}{\shortstack{Major:3\\Minor:50}} & 2:1:1\\ \cline{1-1} \cline{3-3} \cline{4-4} \cline{6-6}
1-2 & & 1:2:1 & 2-2 & & 1:2:1\\ \cline{1-1} \cline{3-3} \cline{4-4} \cline{6-6}
1-3 & & 1:1:2 & 2-3 & & 1:1:2 \\ \cline{1-1} \cline{3-3} \cline{4-4} \cline{6-6}
1-4 & & 4:2:1 & 2-4 & & 4:2:1 \\ \hline
\end{tabular}
\label{table:feedratio}
\end{center}
\end{table}

\section{Proposed method}
\label{sec:Proposedmethod}
The simulation system mainly focuses two parts, `Pheromone Evaporation and Its Diffusion' and `Probability for leaving from chain'.

\subsection{Pheromone Evaporation and Its Diffusion}
\label{sec:pheromoneevaporation}
As for the former part, Pheromone Evaporation and Its Diffusion, the method in \cite{Douzono12} assumed the improper rate of evaporation and diffusion of pheromone in the agent and its behavior. The parameter setting causes a bias for the flock. That is, there is much pheromone in simulation environment partially. The situation increases the agents swarming around them. As a result, it becomes easy to enter Altruism State and two or more bridges are built without the shortest path from a food source to the nest.

\begin{figure}[!tb]
\begin{center}
\includegraphics[scale=0.65]{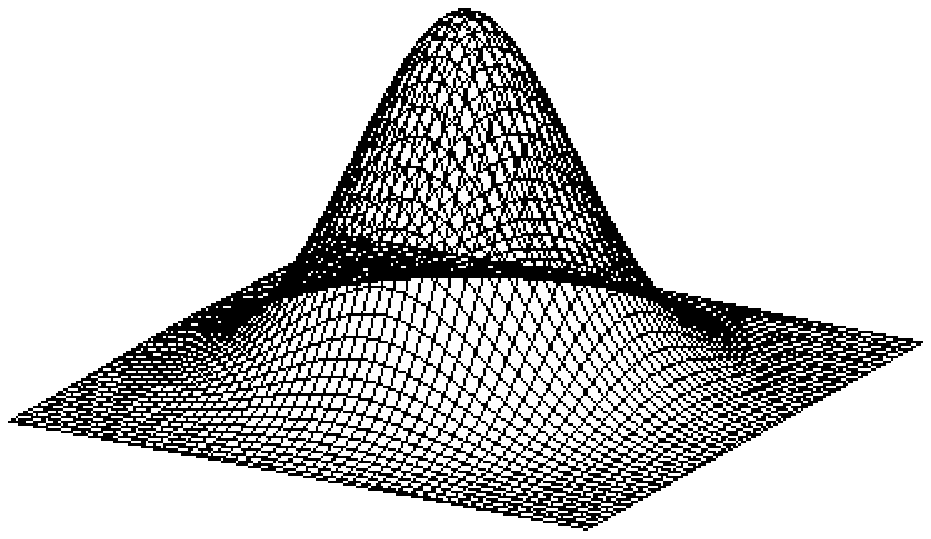}
\caption{The distribution of pheromone}
\label{fig:PheromoneDistribution}
\vspace{-3mm}
\end{center}
\end{figure}

\begin{figure}[!tb]
\begin{center}
\includegraphics[scale=1]{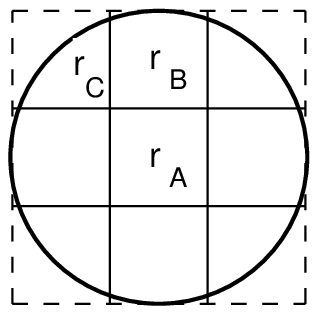}
\caption{The evaporation rate of pheromone}
\label{fig:Pheromoneevaporation}
\vspace{-3mm}
\end{center}
\end{figure}

In order to avoid such a situation, the ratio of pheromone evaporation is defined based on the normal distribution probability as shown in Fig.\ref{fig:PheromoneDistribution}. Fig.\ref{fig:Pheromoneevaporation} shows the transverse plane of Fig.\ref{fig:PheromoneDistribution}. The rate of pheromone in 3D space is set to `$r_{A}$':`$r_{B}$':`$r_{C}$'=0.788:0.043:0.010 in Fig.\ref{fig:Pheromoneevaporation}.

\subsection{The model of Army Ant}
\label{sec:LeavingChainProbability}
The probabilities of an ant entering a chain($P_{e}$) or leaving a chain ($P_{l}$) are depending on the size of the chain. The chain is a small part of constructing bridge. Lioni {\it et al.}\cite{Lioni01, Lioni04} defined these probabilities as follows.
\begin{eqnarray}
P_{e}=C_{e0}+\frac{C_{e1}X_{i}}{C_{e2}+X_{i}},\\
P_{l}=C_{s0}+\frac{C_{s1}}{C_{s2}+X_{i}^{\nu}},
\label{eq:probability_chain}
\end{eqnarray}
where $X_{i}$ is the number of ants in the chain $i$. $C_{e0}$, $C_{e1}$, and $C_{e2}$ are parameters for entering the chain. $C_{s0}$, $C_{s1}$, and $C_{s2}$ are parameters for leaving the chain.

The function $P_{e}$ expresses the idea that the probability for an ant to join the chain grows the number of nest mates already presented and reaches a plateau value equal to $C_{e0}+C_{e1}$. $C_{e0}$ is the value of spontaneous hanging when $X_{i} = 0$. The function $P_{l}$ expresses the probability for an ant to leave the chain decreases with $X_{i}$. 

The ant in the chain does not always stay in the same chain. A certain probability for leaving from the chain is required to realize Altruism Status. Due to interaction between ants, the probability decreases with the number of con-specifics in the chain. The phenomena is ruled by empirical equation very similar to that for Oecophylla\cite{Deneubourg02}. The probability for leaving from chain is given by Eq.(\ref{eq:probability_Leaving}).
\begin{equation}
P_{i}=\frac{a}{1+bX_{i}^{2}},
\label{eq:probability_Leaving}
\end{equation}
where $X_{i}$ is the number of ants in the chain $i$. $a$ is the probability of leaving a chain under a disregard for the number of other agents. $b$ is the parameter of depending the amount of pheromone in chain $i$: $b=\min\{\eta (\log(space_{(x,y)}+1))+\epsilon,1\}$. A theoretical model suggests that these basic mechanisms account for the clustering of insects\cite{Deneubourg02,Lioni01, Lioni04}.

\begin{figure}[!tb]
\begin{center}
\includegraphics[scale=0.3]{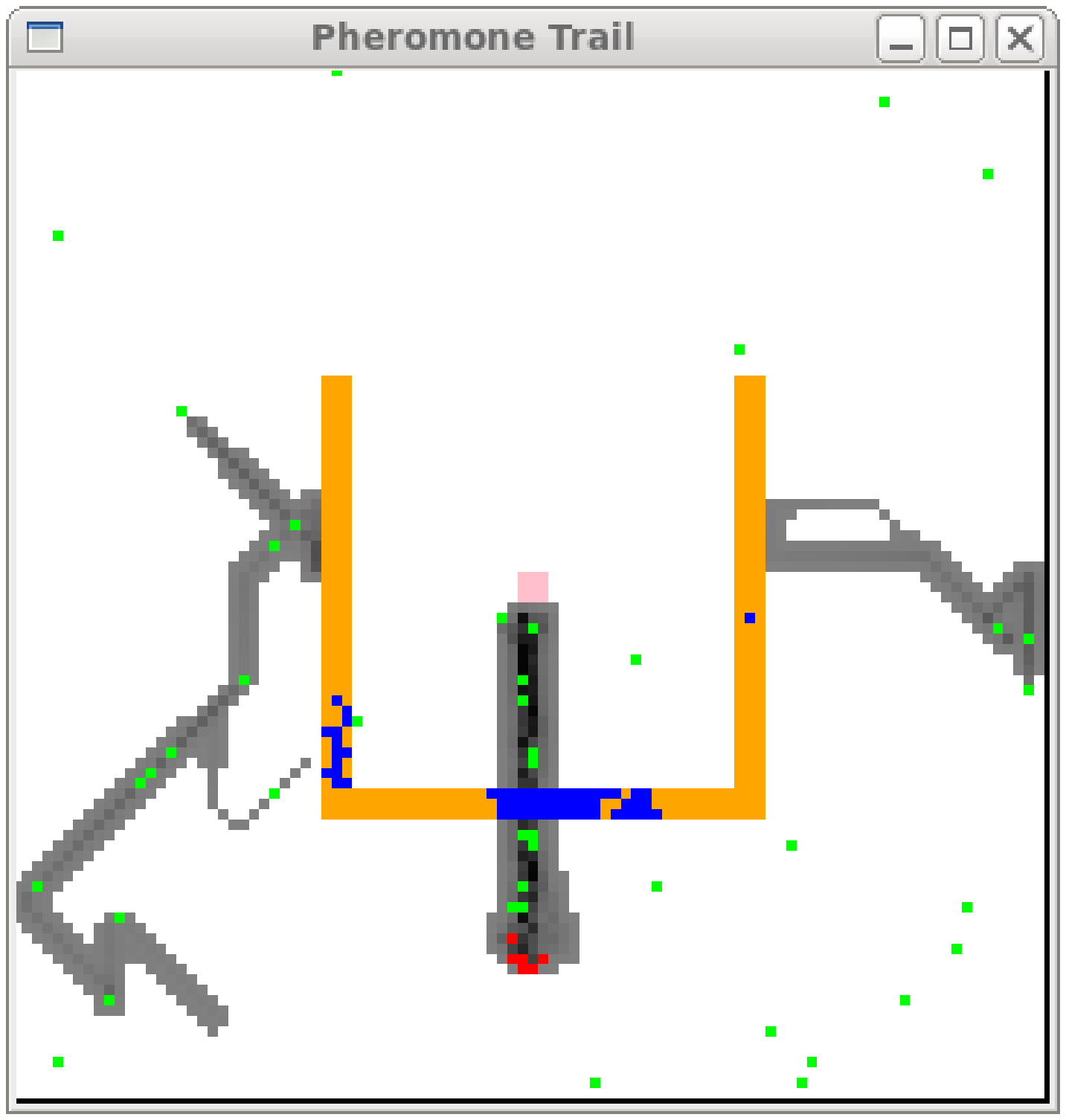}
\caption{The discover of shortest path}
\label{fig:shortestpath}
\vspace{-3mm}
\end{center}
\end{figure}

Ichimura {\it et al.} \cite{Ichimura12} report the agents in altruism situation perform the shortest path by construction of the bridge. Fig. \ref{fig:shortestpath} show the constructed bridge on the way from the feeding spot to the nest. In this paper, we investigate the altruism situation by using the condition in section \ref{sec:AltruismState} for entering a chain and Eq.(\ref{eq:probability_Leaving}) for leaving the chain.

\section{Experimental Results}
\label{sec:ExperimentalResults}
The behavior of army ants at each environment as shown in Table \ref{table:feedratio} was observed. Parameter settings are given as follows by preliminary experiments: $n=2$, $a=0.4$, and $r_{e}=0.05$. At each environment in this paper, 10 trials for each set were executed and the behavior of ants were recorded as the motion video. Each trial is continued till all food in the feeding spots eats up. There are 2 kinds of ants at each environment, Major ants to make a random search and Minor ants to follow the scattered pheromone. In this paper, for almost trials we can observe the following simulation results. During the initial phase as the search of an area for prey, the movement of Major agent will be a key in the change of course to search the subspace, because the Major agent scatters the pheromone while moving in a space. After the discovery of food, the Minor agent catches the food and scatters the pheromone on the way from the spot to the nest. The path becomes congested since there is an obstacle of a ditch. Such situation causes the construction of the bridge on the ditch, since the ants search the shortest path from the feeding spot to the nest. Moreover, the agent swarming around the food increases with the size of food, because more pheromone is scattered while the agent brings a food to the nest. That is, the larger the food in the environment has, the more pheromone the agents scattered.

\begin{figure}[!hbtp]
\begin{center}
\subfigure[A:B:C=2:1:1]{
\includegraphics[scale=0.15]{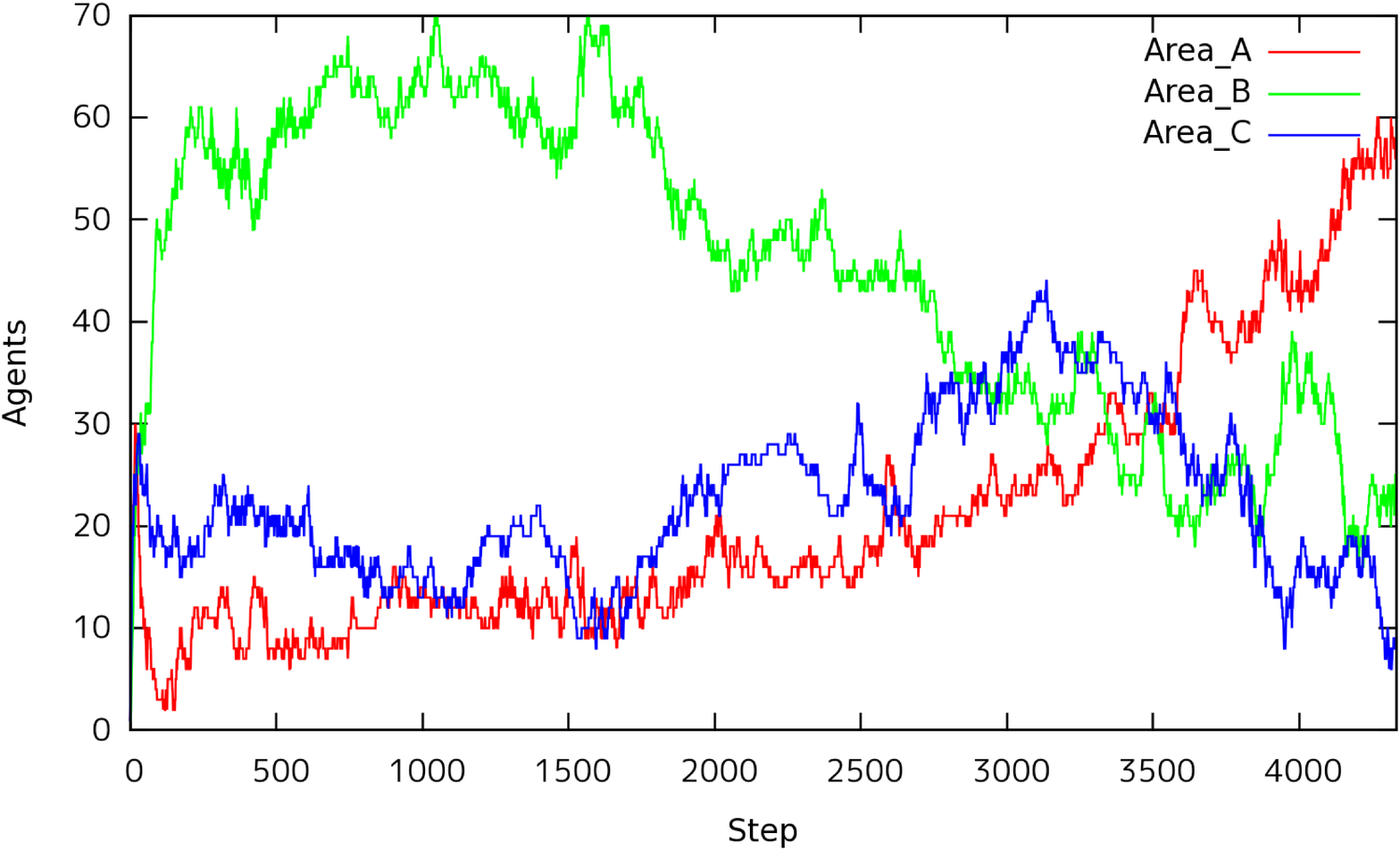}
\label{fig:MultiFeed_Agent100_1_1}
}
\subfigure[A:B:C=1:2:1]{
\includegraphics[scale=0.15]{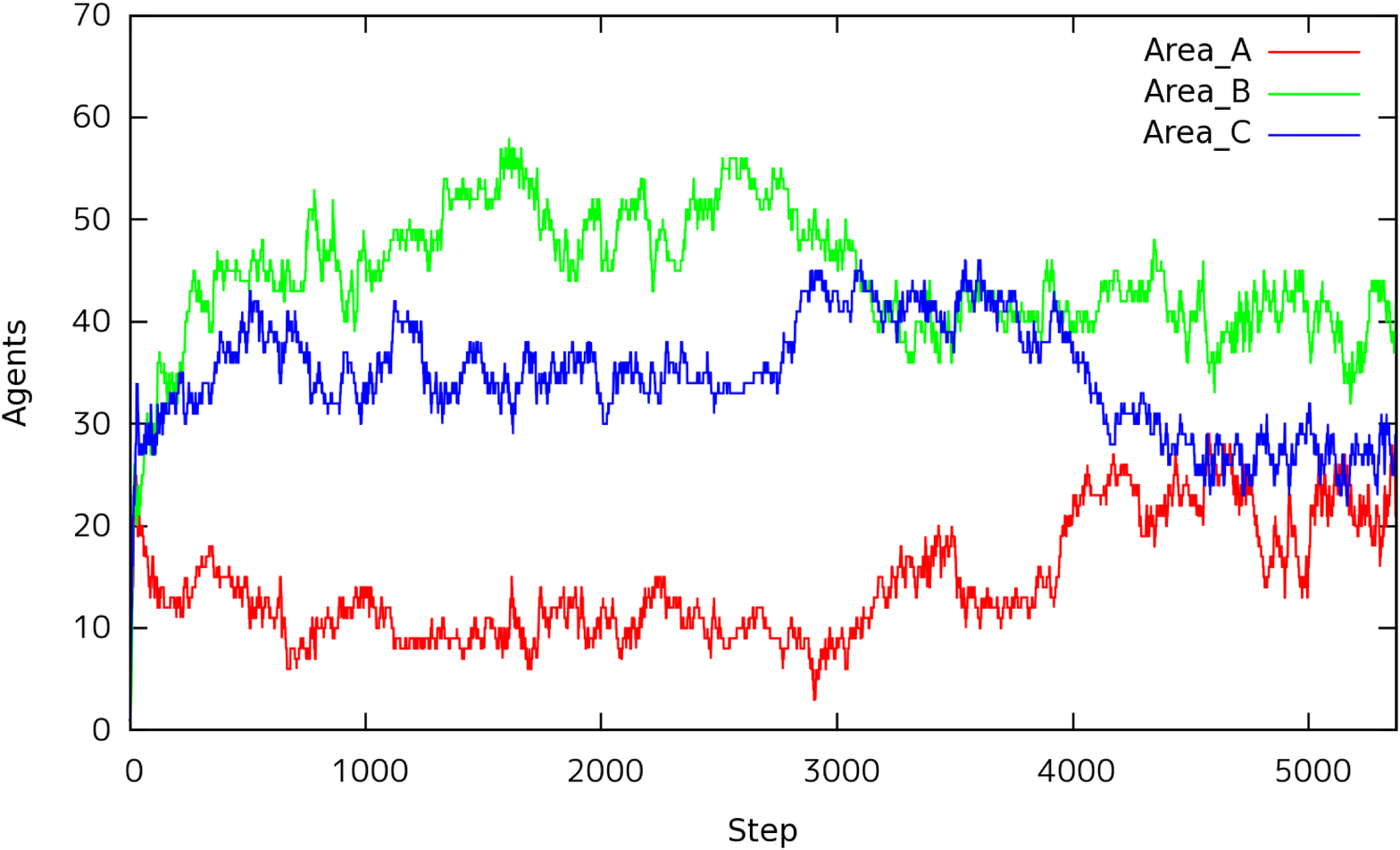}
\label{fig:MultiFeed_Agent100_1_2}
}
\subfigure[A:B:C=1:1:2]{
\includegraphics[scale=0.15]{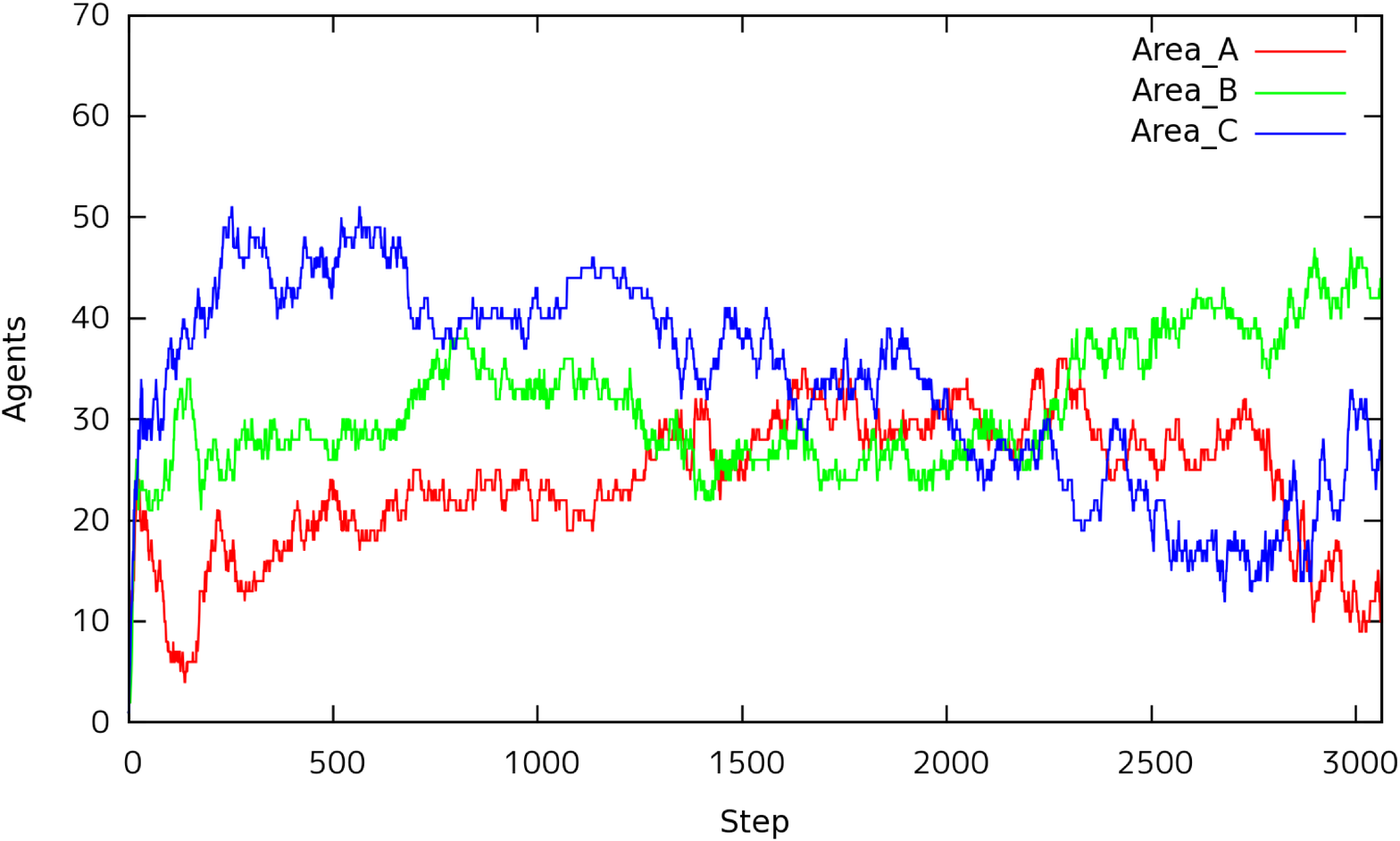}
\label{fig:MultiFeed_Agent100_1_3}
}
\subfigure[A:B:C=4:2:1]{
\includegraphics[scale=0.15]{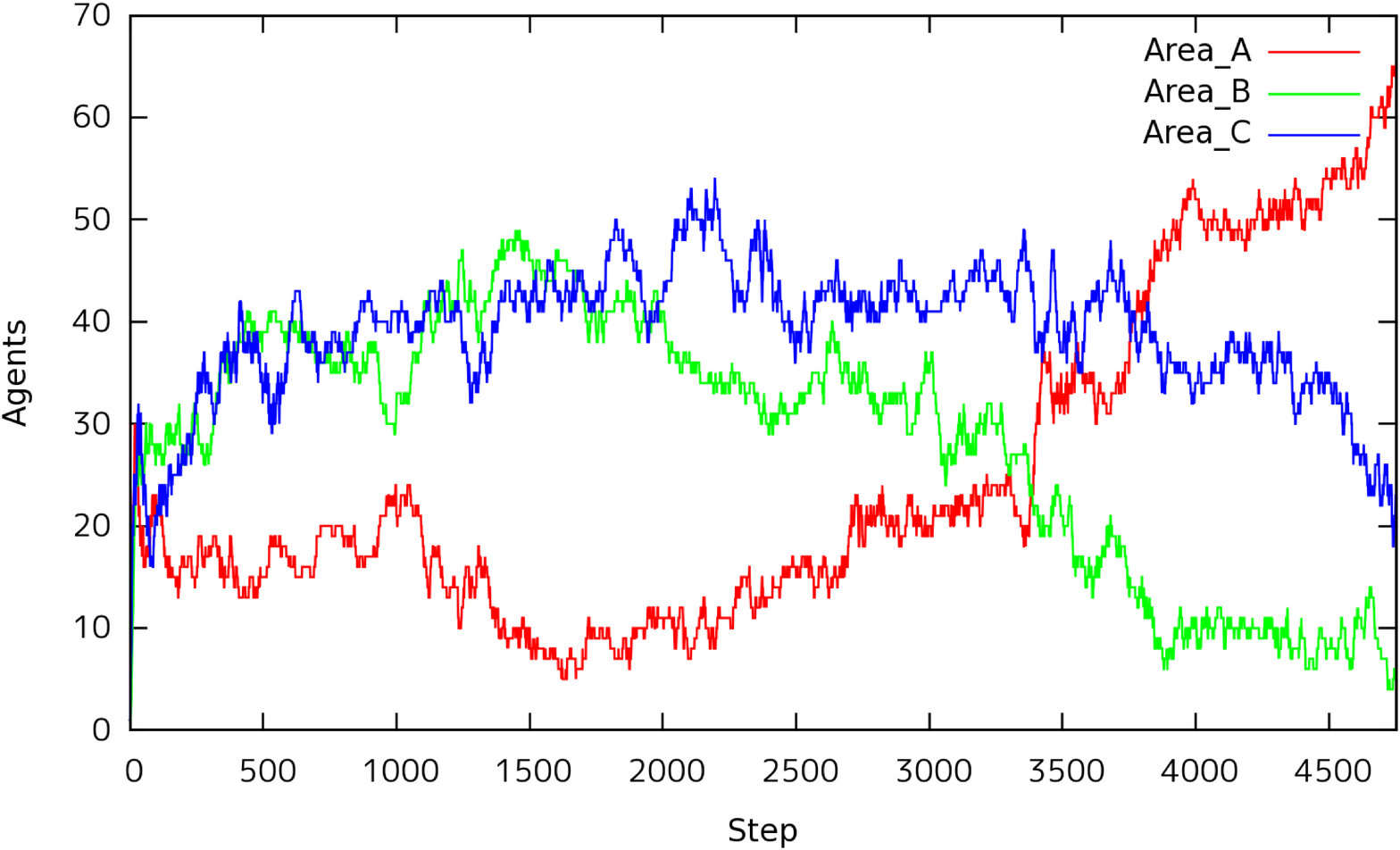}
\label{fig:MultiFeed_Agent100_1_4}
}
\caption{Transition of 100 Agents}
\label{fig:MultiFeed_Agent100}
\end{center}
\end{figure}

Figs. \ref{fig:MultiFeed_Agent100} show the number of agents in each area, A, B, and C, with 100 agents in the environment. In case of 100 agents are in the environment, they are divided into some groups and each group can search the area, respectively. Fig. \ref{fig:MultiFeed_Agent100_1_1}, Fig. \ref{fig:MultiFeed_Agent100_1_2}, and Fig. \ref{fig:MultiFeed_Agent100_1_4} show the transition of agents with $A:B:C=\{2:1:1,\hspace{1em} 1:2:1,\hspace{1em} 4:2:1\}$, respectively. As shown in these figures, we can observe some characteristic behavior related to the altruism in the search space. Note that even if the food size of $A$ is the largest, the food of $A$ remains until $B$ or $C$ is disappeared. Because ants find $B$ or $C$ and bring all food from $B$ or $C$, respectively. On the contrary, Fig. \ref{fig:MultiFeed_Agent100_1_3} ($A:B:C=1:1:2$) shows that the result is beyond our expectations. In the experiment, more agents gather to the area $B$ than the area $A$ at the initial phase, but it took longer time than the area $A$ until they finish carrying the food. Because each simulation was recorded as a motion video, we investigated the detailed behavior of ants. In the almost cases, the constructed bridge was not in the shortest path on the way to the nest, the bridge in the lower area of $B$ was constructed. For this reason, the remaining ant without the participation to the bridge took a roundabout route to avoid the ditch and to search of the shortest path in the area $B$.

\begin{figure}[!hbtp]
\begin{center}
\subfigure[A:B:C=2:1:1]{
\includegraphics[scale=0.15]{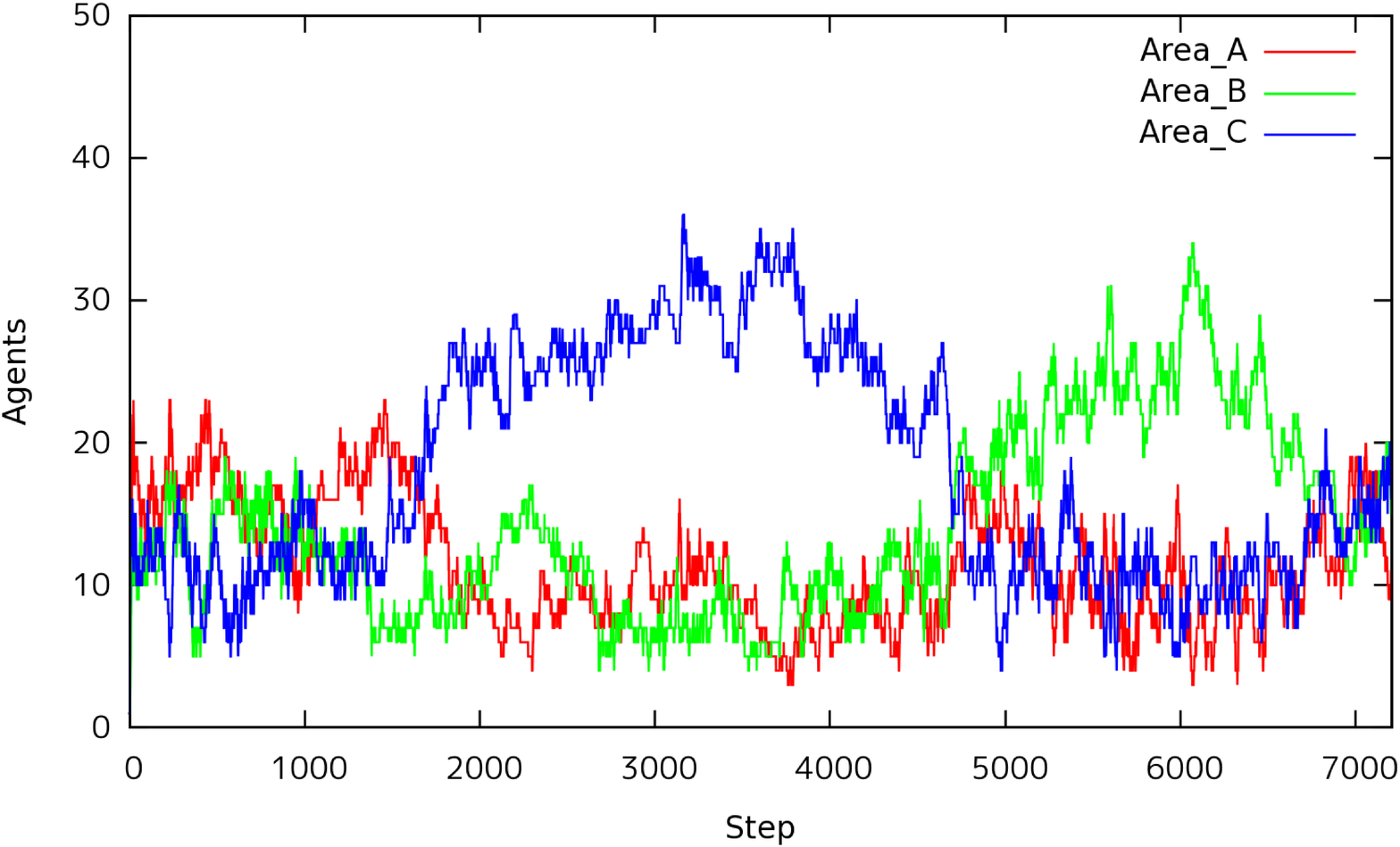}
\label{fig:MultiFeed_Agent50_2_1}
}
\subfigure[A:B:C=1:2:1]{
\includegraphics[scale=0.15]{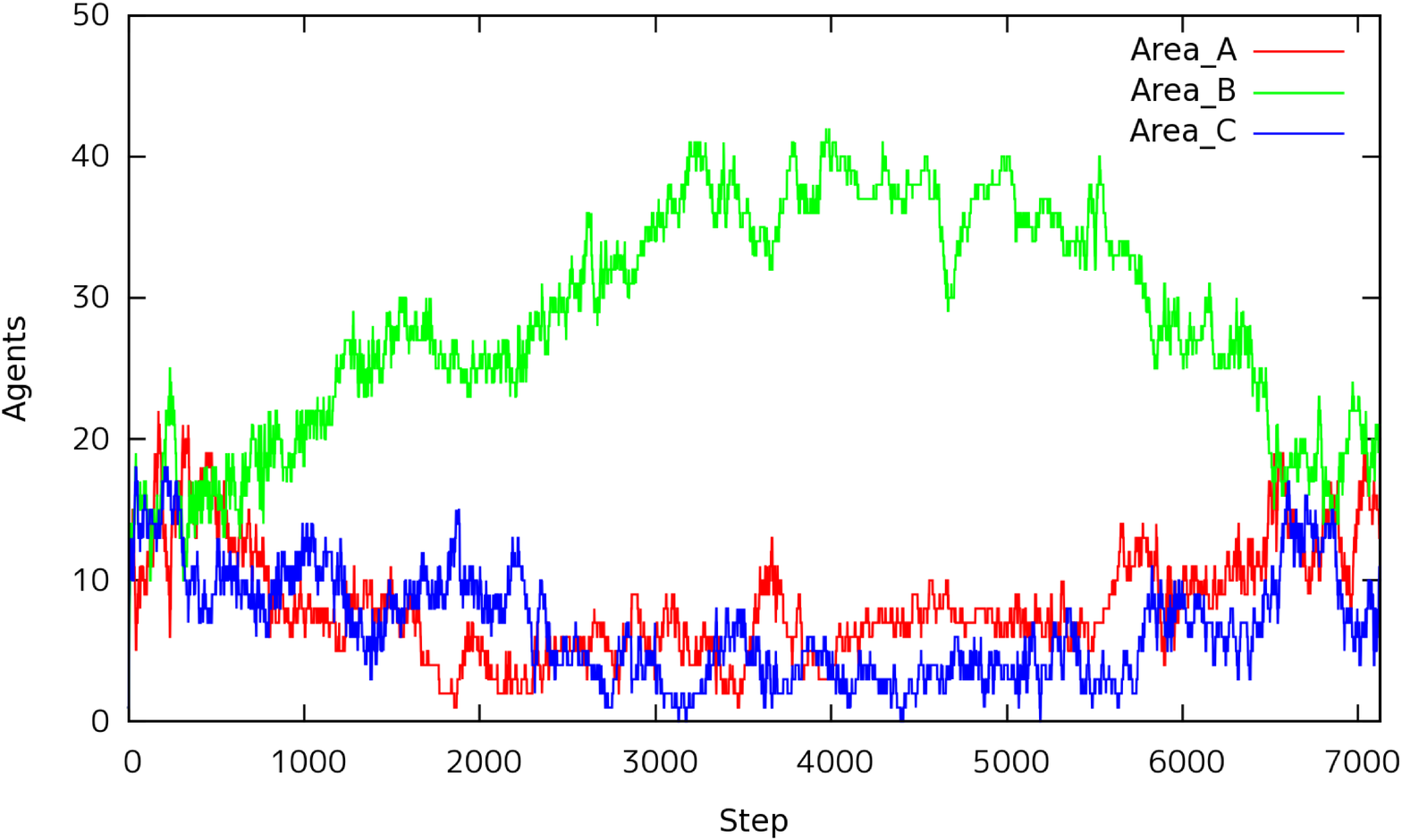}
\label{fig:MultiFeed_Agent50_2_2}
}
\subfigure[A:B:C=1:1:2]{
\includegraphics[scale=0.15]{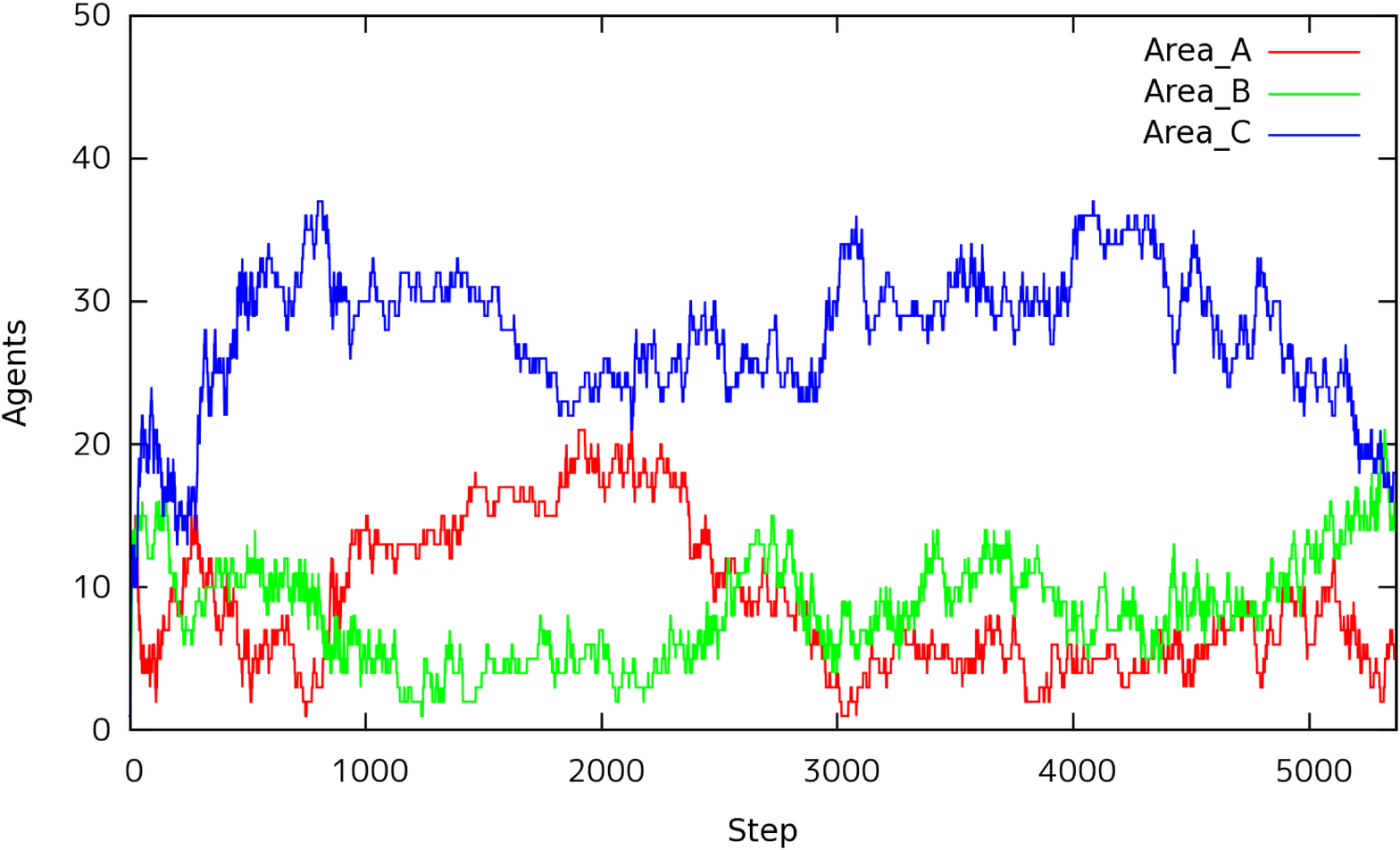}
\label{fig:MultiFeed_Agent50_2_3}
}
\subfigure[A:B:C=4:2:1]{
\includegraphics[scale=0.15]{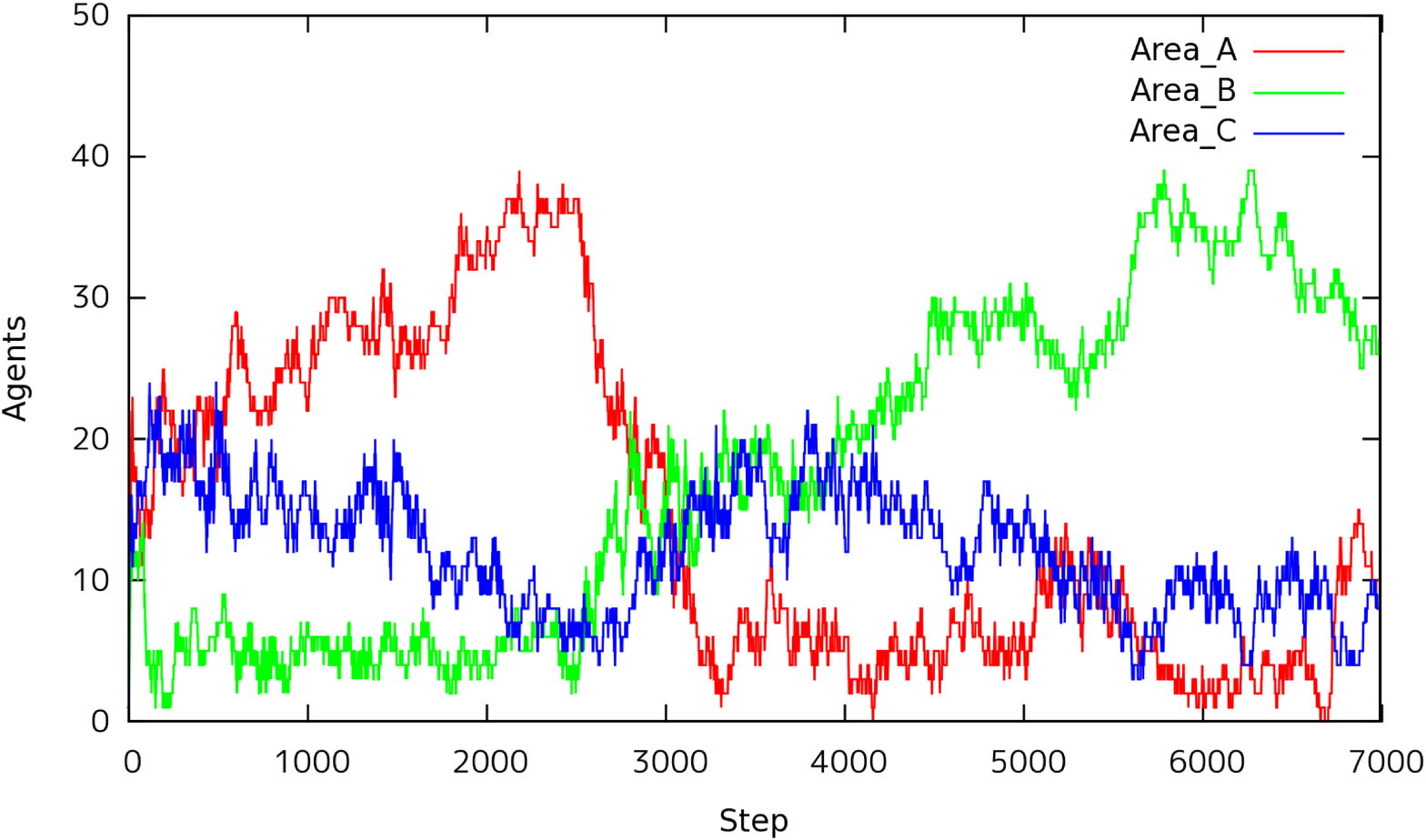}
\label{fig:MultiFeed_Agent50_2_4}
}
\caption{Transition of 50 Agents}
\label{fig:MultiFeed_Agent50}
\end{center}
\end{figure}

Figs. \ref{fig:MultiFeed_Agent50} also show the transition of agents under 50 agents in the environment. The situation has 50 agents in the environment, they cannot search the space sufficiently. The ants did not divide into some search group and then the search of area was processed sequentially such as $A \rightarrow B \rightarrow C$. Fig. \ref{fig:MultiFeed_Agent50_2_1} and Fig. \ref{fig:MultiFeed_Agent50_2_4} show the transition of agents in case of $A:B:C=\{2:1:1,\hspace{1em} 4:2:1\}$, respectively. As shown in these figures, we can observe some characteristic behavior related to the altruism in the search space. That is, the feeding spots are disappeared in the order of larger spot. On the contrary, Fig. \ref{fig:MultiFeed_Agent50_2_2} ($A:B:C=1:2:1$) and Fig. \ref{fig:MultiFeed_Agent50_2_3} ($A:B:C=1:1:2$) show that the result is beyond our expectations. In case of Fig. \ref{fig:MultiFeed_Agent50_2_2}, the constructed bridges were not in the shortest path on the way to the nest as shown in Fig. \ref{fig:MultiFeed_Agent100_1_3}. In case of Fig. \ref{fig:MultiFeed_Agent50_2_3}, it is an interesting case, and almost agents will make bridges place to place in the ditch from area $C$ to the nest. The remaining agents should deliver the food to the nest, however very few agents cannot take all of them. The scattered pheromone was smaller than the evaporated pheromone. Therefore, the pheromone around the bridge disappears and then the bridge was destroyed, because the agents in altruism situation depart the bridge in the condition of less pheromone. The agent leaving from the bridge moves to area $A$, but there are only a few agents in the area. From such results, only few agents in the environment with large size of food cannot get the altruism situation easily.

\section{Conclusive discussion}
\label{sec:Conclusivediscussion}
We developed the army ant inspired social evolutionary system which can perform the altruism. There are 2 kinds of ant agents communicated with each other via pheromones. Moreover, the pheromones evaporate with the certain ratio and diffused into the space of neighbors stochastically. In order to avoid the over-concentration in the chain, the probability of leaving from a chain is introduced. The system with the facilities can find the optimal place of bridge. The path through the bridge is the shortest from foods to the nest. In this paper, the behaviors of ant under the environment with multi feeding spots and the adequate number of agents were observed. The altruism behavior in the few agents to the size of food spot is hard to keep its situation. Such observations of behaviors in the computer simulation strongly will contribute to the shift to knowledge and power from the individual to the collective. We will develop the autonomous intelligent robots with the altruism behavior and investigate the collective intelligence system in near future.

\end{document}